# Computers and the Conservation of Energy

Grenville J. Croll

*grenville@spreadsheetrisks.com*

*The purpose of this report is to show that computer and allied technologies can be used to increase energy efficiency. The report is divided into transport, industrial, commercial and domestic sections which correspond to the major energy consuming sectors of the economy. Each section considers the various ways in which energy can be saved by the use of the computer. The report concludes that it is economic to incorporate computer based energy management systems in a wide variety of applications and that it is important that this capability is realised on a large scale. A comprehensive reference list and a bibliography are included.*

**CONTENTS**

## 1.0 INTRODUCTION

Recent events in the oil supply industry have highlighted the need for conservation, since there is no longer the certainty of supply that there has been in the past. The disruption in Middle Eastern oil supplies has done more than to cause a short term disruption: it has shown that problems of supply are likely in years ahead unless action is taken now to develop alternative sources of supply.

Alternative energy sources share a number of characteristics. Compared with conventional oil, they generally require higher investment costs, longer development times and may have a greater environmental impact. Moreover, these alternative energy sources will have to be developed in a climate of uncertainty because there is no guarantee as to when the anticipated demand for them might develop. Against this background, energy conservation, with its inherent advantages of being indigenous and clean, is increasingly being considered as a substitute for energy supply in its own right.

In order to maintain the present standard of living, it is necessary to conserve energy by using it more efficiently rather than by prohibiting its use in certain situations. The latter course of action is reserved for times of crisis, the former for more normal times.

The computer is an ideal tool with which to achieve the desired increase in energy efficiency because of its ability to process large amounts of information quickly and cheaply. Computers have been used in energy conservation applications in the past. They are now being more widely used because of their decreasing size and cost for a given processing power.

The purpose of this report is to show that the computer has considerable potential as a means for conserving energy by describing in some detail current applications.

### 1.1 Computer Applications

The graph on the following page shows how much of the UK's primary energy is consumed by each category of user. Correspondingly, this report is divided into transport, industrial, commercial and domestic sections in order to reflect the structure of the graph, and to provide a means of grouping together common problems and solutions. The report describes individual application areas of energy conserving computer technology within each of these sections.

The transport sector can benefit from improved route planning, vehicle scheduling and urban traffic control through the analysis of the complex logistical problems involved. Microprocessors are being used in ignition, carburetor and fuel injection systems to improve the efficiency of internal combustion engines. Industry can benefit from the more widespread implementation of process control systems and by better process design.

The commercial and domestic sectors benefit from the increased use of computer based heating, ventilation and air-conditioning equipment.

The extent to which the computer may be used to conserve energy within each of the four sectors depends to a large extent upon the cost-effectiveness of the computer, the flexibility of choice of fuel within the sector and the number of end uses to which the computer may be applied.

It is evident that the transport sector is the most susceptible to computer usage since the choice of fuel is restricted and conservation depends to some extent upon the marginal refinement of existing technology. Also, there is a wide range of problems to which the computer may be applied

**Figure One - Primary Energy Usage by Sector \***

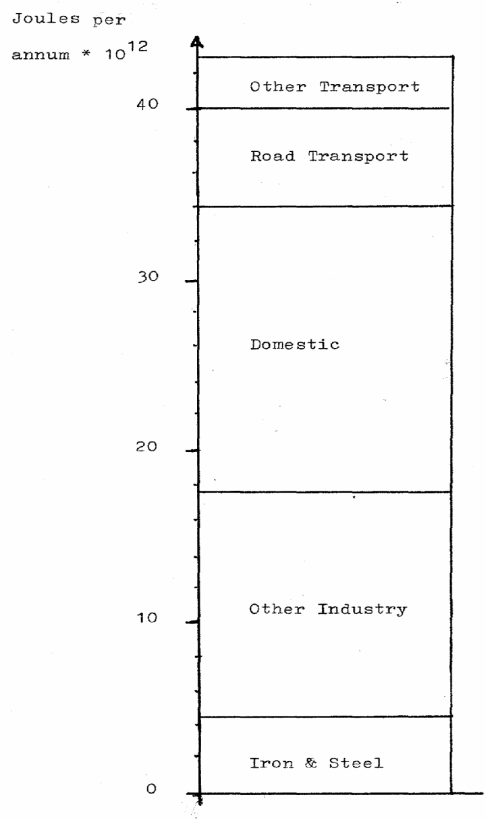

*\* Actual primary energy consumption excluding non-energy usage. Transformation losses have been allocated to the various end uses.*

The commercial and domestic sectors are less well suited to the use of the computer because there are very few end uses of energy in these environments. Energy is mainly used for heating, cooking and lighting. It is certainly much more cost-effective to save energy in the home and office by insulating the loft, walls and windows than it is to install a computer based energy management system. However, at some future time when there is little to be gained by the installation of additional insulation, it may become cost-effective to install a domestic central heating controller. That is not to say that energy management systems will not be used in the home in the near future, on the contrary, they may be installed for convenience, aesthetic or status reasons well before they are cost-effective.

The industrial sector is well suited to the use of the computer for energy conservation purposes. However, the large number of possible end uses can be categorised into a small number of general application areas. Furthermore, the intrinsic energy intensity of some of industry will have ensured that a lot of attention will have already been paid to energy conservation so that further efforts are not justified.

Of the large number of possible applications of the computer in the field of energy conservation, it is apparent that only a small number will be responsible for the greatest overall saving. The importance of this small number should not be allowed to obscure other applications since, on a national or global scale, every saving is of the utmost importance.

## 2.0 COMPUTERS, CONSERVATION AND TRANSPORT

The energy consumed by the transport sector amounts to 23% of the total UK domestic consumption [88]. The importance of conservation in the transport sector, although great already is increased by the fact that the sector relies almost exclusively on petroleum products for its motive power. Increased fuel efficiency in transport will release much needed petroleum for other consumers who rely upon it as a basic chemical feedstock and will provide extra time in which to develop alternative fuels.

Computers can be used in four distinct ways within the transport sector as an aid to energy conservation.

Firstly, they can be used in the planning of transport systems. The cost and length of time necessary to complete a planning study is usually large, but so too are the benefits to be gained. A transport system, once established lasts for many years and it is important that it provides an energy efficient and low pollution solution to the original planning problem.

Secondly, computers may be used in the scheduling of vehicles of all types. The number of miles a goods vehicle travels during the course of a day is dependent entirely upon the skill of the route planner in selecting the shortest route out of all those possible. A human route planner can not be expected to choose the optimum route because of the inevitable complexity of the problem. To the computer, the complexity is not quite so relevant, and in many circumstances it can be relied upon to produce a better solution.

Thirdly, by virtue of their small size and relatively large computing power, microprocessors are now being used in vehicle engines and transmission systems. Microprocessors are able to control accurately the critical factors relating to engines and transmissions with a view to increasing efficiency and decreasing pollution.

Finally, computers and telecommunications services are now being recognised as possible substitutes for transport in the first place. Detailed studies have shown that certain types of everyday travel, particularly business travel, may be capable of being replaced by some form of computer facility.

The following sections describe in detail the realities and possibilities of conservation in the transport sector through the increased use of computers.

### 2.1 Engine Management

The timing and carburetor systems on most present day European cars are of a strictly mechanical nature. They are made of precision components and have to be accurately adjusted to ensure optimum vehicle performance. Being of a mechanical nature, these systems are prone to wear and gradually slip out of adjustment with time, so reducing engine efficiency. It is then desirable for the engine to be retuned. This rarely happens immediately and so the engine is allowed to run below optimum efficiency for some time. Ways in which this wastage can be reduced are therefore highly desirable. Vehicle manufacturers are constantly striving to provide the maximum function and reliability in vehicle components for the least cost. Until the early seventies, it was not worthwhile for manufacturers to drastically alter the basic conceptual designs of the engines they produced since the current ideas worked perfectly well and provided a reasonable fuel consumption. At about this time, with the increasing price of petroleum, it became economic to at least consider investigating electronic ignition systems, a major break from tradition. The first devices were very crude by today's standards and often consisted of little else but a power transistor, the base of which was driven by the contact breaker [20]. Early systems such as this would have required much improvement before they could have been mass-produced. The next generation of devices were based upon the capacitor discharge principle. These systems provided a very high spark voltage and so ensured better ignition, partly because dirty spark plugs did not degrade the spark to a great extent, Better ignition, and a correspondingly better burning of the combustion mixture meant that there was a very slight reduction in fuel consumption. Capacitor discharge systems became widely available on the accessories market and some up-market cars were produced with them as a standard fitting.

The large fuel price increases of the mid-seventies spurred on interest in the development of electronic ignition, and manufacturers moved on from analogue to digital systems. These were initially of the hard-wired type and suffered badly from problems due to the large number of soldered joints and high power dissipation. The advantages of digital systems were that they could process more information relating to engine operating parameters, and were very much more independent of environmental influences.

Manufacturers' attention is now focused upon microprocessor based systems because they offer the ultimate in capability, size, reliability, flexibility and cost. The Americans and Japanese are well ahead in their design and development programs of these systems, partly because they have to meet stringent fuel consumption and emissions control regulations (see Table One).

The use of microprocessors in automobiles is now diversifying to include the control of fuel injection and carburetor systems, and in isolated instances, the control of vehicle transmission chains.

The emerging concept is now that of total engine management by microprocessor, where each facet of the engine's operation is controlled by one or more microprocessors. This concept may be developed even further such that total vehicle management by microprocessor will become a reality in the not too distant future.

**Table One - Emission and Fuel Economy Requirements in the USA[28]**

| Year | HC/C0/NOx* | CAFE (mpg) |
|------|------------|------------|
| 1978 | 1.50/15/2.0 | 18 |
| 1979 | " | 19 |
| 1980 | 0.41/7.0/2.0 | 20 |
| 1981 | 0.41/3.4/1.0 | 22 |
| 1982 | " | 24 |
| 1983 | " | 26 |
| 1984 | " | 27 |
| 1985 | " | 27.5 |

HC - Unburnt hydrocarbons, CO - Carbon monoxide, NOx - Oxides of Nitrogen (* Units not given), CAFE - Corporate Average Fuel Economy

2.1.1 The need for more precise engine management

Figures Two and Three show the relationship between the air/fuel ratio and the concentration of pollutants in an IC engine and the relationship between the air/fuel ratio, power output and fuel consumption. The region labeled stochiometric is that ratio of fuel and air at which perfect combustion can take place. The purpose of more refined engine management systems is to keep the fuel/air ratio slightly on the lean side of stochiometric without prejudicing power output, fuel consumption or pollutant level. This level of control is difficult to achieve because of the many conflicting requirements.

The situation is complicated further in the USA because of the use of three-way catalyst converters on the exhaust system to remove pollutants, and the use of exhaust gas recirculation (EGR) for the same purpose. Both of these systems require monitoring and control, and the various parameters that affect these systems also affect the engine. It is apparent therefore, that control of an automobile engine is complex, and that the microprocessor is likely to be the only satisfactory method of achieving this degree of control.

**Figure Two – Power & Fuel consumption variation by Air/Fuel ratio**

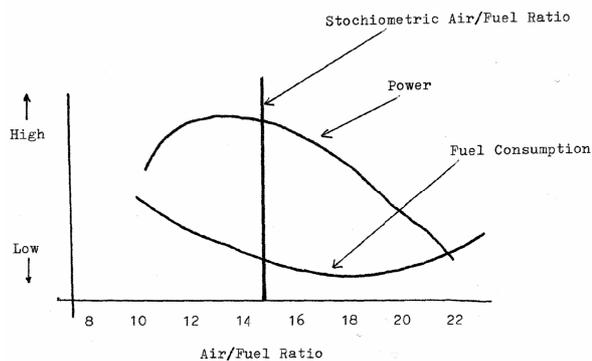

**Figure Three – Exhaust Emissions variation by Air/Fuel ratio**

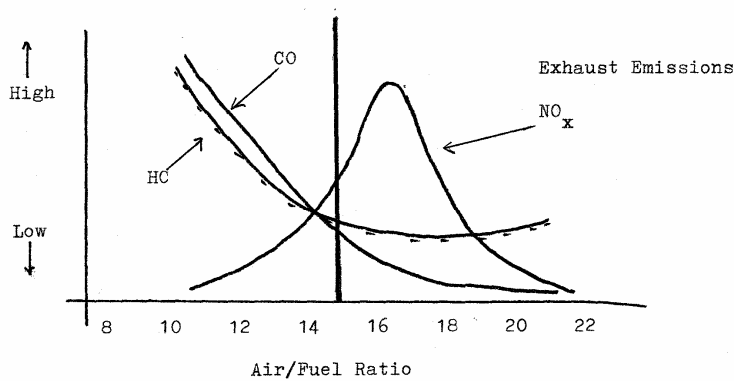

2.1.2 Reliability

The reliability of on-board vehicle electronics is of the utmost importance. It is essential that units function for long periods of time before failing or falling out of adjustment.

Unless vehicles fitted with advanced engine management systems are seen to be reliable by the general public, then they will be rejected out of hand even though the devices bring about significant reductions in fuel consumption.

The automobile environment is extremely harsh and is particularly hostile to electronic systems because of the presence of electromagnetic interference, high temperatures, voltage transients and other phenomena.

Increasing levels of integration are helping to solve the reliability problem. A higher level of component integration means that fewer electronic components are exposed to the environment. As a result, the environment does not affect the circuitry to the same extent. Reliability is being increased further through the development of logic families more suited to the hostile automobile environment.

The question of reliability in automobile electronics is well covered in the literature [30][37] which describes problems encountered and contemporary solutions. Continued development of engine management systems from the component level upwards should result in systems that are guaranteed for life.

2.1.3 Electronic Ignition Systems

The Chrysler Corporation of the USA introduced in the 1978 model year, a microprocessor based ignition system [31]. The system comprised an RCA CMOS microprocessor incorporating 1024 bytes of ROM, a 32 byte RAM and a custom I/O circuit. The system was installed in 30,000 cars to determine the production requirements for a high volume design. The design was successful and production was extended into the 1979 model year, in larger quantities. Similar systems are available in production quantities from most of the major American, European and Japanese manufacturers [20][27][28][36][38][40]. There cannot therefore, be any doubt about the feasibility or economics of microprocessor based ignition systems, or the willingness of manufacturers to develop them.

It is possible to implement ignition systems in other than microprocessor technologies, however, none have the inherent capability for expansion and refinement. The control of the ignition in an IC engine is not an easy task, and many engine parameters have to be monitored to ensure that the engine is functioning optimally. Below is a selection of the environmental and engine parameters that are monitored in some manufacturers systems:

    - Coolant temperature
    - Air charge temperature
    - Ambient Temperature
    - Absolute ambient air pressure
    - Inlet manifold air pressure
    - Engine RPM
    - Crankshaft position
    - Throttle position

From these inputs it is possible to determine the exact instant at which a pulse should be sent to the power stages of the system, and hence to the spark plug. Since no high voltage current switching semiconductors are available (i.e. 30,000V), the ignition coils of systems such as these have multiple secondary windings so allowing the current to be switched at low voltages.

The development of microprocessor based ignition systems has depended to a large extent upon the availability of appropriate sensors [26][33]. The EGO sensor is a new development and is constructed out of zirconium dioxide. The substance has a non—linear behavior- the voltage it produces (since it acts as a battery) changes very rapidly when there is no oxygen in the exhaust gases, indicating complete combustion. The crankshaft position sensor, also a new development, utilises the wiegand effect and is generally much more accurate than simpler electromagnetic devices. Semiconductor pressure sensors are also available, and have been specially developed for automobile applications. Position and temperature sensors use existing resistor and thermistor based technology.

Research and development of advanced ignition systems is continuing at a rapid pace. It is to be expected that development will continue to the point where the ignition system becomes part of a higher level engine management system.

2.1.4 Fuel Injection and Carburetor Systems

Fuel injection and carburetor systems differ widely in their method of operation. Both systems are very much amenable to some form of microprocessor control. Detailed descriptions of some manufacturers systems are provided in the literature [18][21][22][34].

Fuel injection permits the precise metering of the quantity of fuel required by the engine according to the operating conditions. The quantity of fuel injected can be optimised by means of various sensors, actuators and a microprocessor. This fuel quantity is injected continuously or intermittently directly in front of the engine air intake valves. This results in a whole series of advantages compared with a conventional carburetor: A precise fuel quantity is injected under all operating conditions; There is no transport time from the metering point to the intake valves; There is no wetting of the intake manifold; And finally, there is even distribution of the fuel to all cylinders.

The purpose of a microprocessor control system is therefore to ensure that exactly the right amount of fuel is injected into the engine at the right time. The microprocessor performs this task admirably and a system developed by Lucas [35] gives particularly good idling, cruising and acceleration performance.

Carburetor control systems, as described by Scheile [39], have similar objectives to those of fuel injection systems: that is to provide precise fuel metering under all engine operating conditions. Carburetor systems are however, much cheaper than their fuel injection counterparts, and this accounts for the low penetration of fuel injectors into the market. As fuel and emissions control becomes stricter, fuel injection will become more common since a given engine will perform better from all points of view if fuel injection is fitted.

**2.2 Transmission Control**

Driver behavior determines to a certain extent the fuel consumption of a vehicle. Drivers do not respond well to changes in engine speed, engine load and road gradient: they usually select an inappropriate gear which results in the engine running less efficiently than is possible.

Simulations of driver behavior [15] have revealed the extent of the problem. A TRRL simulation study investigated the effect on fuel consumption of changes in overall gearing and differing models of driver behavior. The results show that both driver behavior and gear ratios have a significant effect on fuel consumption, and that there is a strong case for removing the choice of gear from the driver by the use of an efficient automatic transmission.

Further evidence is provided from analyses of engine characteristics [16]. These studies show that the efficiency of an IC engine is very sensitive to changes in the speed and load of the engine.

A prototype contro1ler has been built at UMIST as a plug compatible replacement for relay based systems installed in heavy vehicles manufactured by the Bus and Truck division of British Leyland. The system is microprocessor based and continuously monitors the status of a variety of analogue and digital inputs. When the road speed changes, or there is a signal from the gear stick, the system checks the validity and safety of the gear change, if one is required, and then engages the appropriate gear. The system has been successful and is in the process of further refinement at UMIST [19].

Although the energy savings possible through the use of the above system are not quoted, a recent report in The Times newspaper [43] revealed that a similar system intended for use in cars is 25% more efficient than conventional automatic transmissions.

Enhancements to automatic transmission systems have been proposed that could result in even greater fuel savings. Schwarzkopf and Leipnik [24] have proposed an algorithm that allows the driver to choose an arbitrary steady state velocity for a level road, but modifies the speed for optimal fuel consumption on various gradients.

Further work upon automatic transmission systems should result in their more widespread introduction, particularly in heavy goods vehicles [32] where there are already so many gears that the choice of the correct one is difficult.

## 2.3 Transport Planning

Computers can be used in the planning of transport systems in order to optimise energy consumption or other factors [1][3]. Typical transport planning problems include the siting of depots, determination of fleet size and the planning of urban traffic systems. The importance of planning systems such as these correctly is due to the fact that the operation may be carried out once only, or very infrequently.

Two techniques are commonly used in planning problems: linear programming and simulation. The former is used where the problem is well understood and a complete set of data can be obtained, the latter when the problem is very complex, and the individual items are neither independent nor fully understood.

### 2.3.1 Depot Siting

The general problem of depot siting consists of locating a depot to which supplies are transported in bulk, preferably by a low cost large batch size media such as barge, ship, pipeline or rail, and from which deliveries are made by road to customers widely dispersed throughout the area. Often the selection of sites is limited and the problem is to predict future supply costs and quantities, delivery costs and quantities, storage and. handling costs, and to select the site at which these will be minimised. The depot siting problem is closely inter-related to the fleet planning problem because delivery quantities must be combined into loads to be carried by specific vehicles. The solution of the depot siting problem may be influenced by vehicle size since this will influence the manner in which delivery points are grouped to form vehicle loads.

A solution to the depot siting problem has been proposed by Holliday and Wren [82]. Their method relies upon the use of a vehicle scheduling program to develop routes for a typical days deliveries for a selection of possible depot locations. The depot site that results in the lowest overall mileage or cost can then be considered to be the best likely location. In practice, there are few occasions when a company can consider a drastic reorganisation of its depots, and the most common problems are in connection with the possible closure or re-siting of a single depot, or the opening of a new depot. In such cases, the number of possible depot sites that may be considered is small because of constraints imposed by other aspects of the company's operations.

2.3.2 Route Planning

Computers have been used in the determination of routes for all modes of transport and Wren [3] gives a very full treatment of the subject. When deciding vehicle routes, many different factors must be taken into consideration such as the nature of passenger demand, desired concentration of service and the desired quality of service. As with the depot siting problem, the planning of transport routes may only be done infrequently. An additional restriction with, say, bus transport is that routes may not be drastically changed because large groups of people may rely on a particular service for social reasons. As an example of the benefits of computer methods from an energy point of view, the Wallasey local authority [11] revised their bus services and returned a mileage saving of 5%.

**2.4 Computers and Vehicle Scheduling**

The vehicle scheduling problem [5] is one of designing routes whereby vehicles at one or more depots cater for the requirements of customers either by the delivery or the collection of goods. Restrictions on the duration or distance of journeys and vehicle capacity may also be met, The problem of vehicle scheduling is a variant of the travelling salesman problem. The problem lies not in its description, which is simple, but in its solution which is a combinatorial problem of immense difficulty of solution.

The desired objective in scheduling vehicles is usually to minimise the total number of vehicles used, or to minimise the total time or distance for the journeys. Minimising the total mileage covered by the fleet of vehicles is a very common objective and has obvious energy conservation implications. There are often many practical constraints imposed upon the scheduler by the very nature of the problem. Typical restrictions include vehicle capacity limitations, drivers legal working hours, priority of customer service, prescribed time for delivery or collection and restrictions on the type of vehicles that can be received. There are also relaxations, or negative constraints, including such allowances as delivery outside normal working hours, delivery with a different type of vehicle and part delivery of orders. These often prove difficult to incorporate in current scheduling systems.

Quite often, the solutions produced by a vehicle scheduling system bear little resemblance to those produced manually, but the savings are nevertheless real. Where customer orders are highly variable and the system very volatile, it is possible to link the scheduling system to other aspects of the companies business, For example, order processing, picking, delivery instructions etc. The schedules can then be produced daily from the variable data. At the other extreme, delivery patterns of some organisations are so constant that they remain unchanged for many months. In these cases it is worthwhile spending more time and money on producing a really good solution at the outset since this will be used for long periods and a measure of the quality of a solution is the length of time spent obtaining it.

### 2.4.1 The Scheduling of Road Transport Fleets

There is no doubt that transport costs will continue to rise, and with the other constraints of delivery restrictions, computer based scheduling systems will be developed to more sophisticated levels and will be used by more organisations.

A survey by a consultancy, as reported by Collin [23] indicates that the physical distribution costs average just under 10% of sales revenue and that the transport costs average 4% of sales revenue of a typical company. The survey also showed that most organisations under-estimate these costs and not one of the companies surveyed used a computer based scheduling package. Eighty one per cent of those surveyed however, used computers for other purposes.

The savings claimed for those who use a scheduling system are dramatic - 10 to 50 per cent of transport costs can be saved. Most package suppliers claim about 25% of transport costs. Up-to-date estimates for the likely benefits to be obtained from the widespread introduction of computerised vehicle scheduling systems are not available. However, the NCC [1] published some figures in 1969. The figures revealed that approximately 1.5% of transport costs could be saved in the UK by the use of vehicle scheduling. Energy costs are, of course, a significant portion of transport costs (a 1968 estimate [23] gave 13-24% of the cost of running a heavy vehicle).

### 2.4.2 The Scheduling of Other Fleets

Vehicle scheduling does not, of course, apply only to road transport fleets. Other modes are equally open to the scheduling of resources by computer methods.

A recent article in The Times [42] gave some details about a new flight planning system. The system is based at Gatwick and is linked to a computer in California that calculates in a few seconds the most fuel efficient route between any two airports in the world, based upon current meteorological data. The designers estimate that of the £1000m that will be spent on aviation fuel in Britain this year, about £200m of that could be saved by this system. The system has been tried by a few operators and one of them reported a 1.5 hour flight time reduction on a flight between London and the West Coast of America.

British Rail (BR) has not been slow to consider the use of computers in scheduling some of their massive operations. Their locomotive scheduling system resulted in a 14% saving in the number of locomotives they needed to use. Their freight business is now completely automated by the TOPS [7] system. The business of moving freight is very complicated and involves the scheduling of 300,000 wagons. They estimated the savings to be of the order of £1m per month.

Computers have been involved in BR timetabling for many years, however, the systems they use are fragmented. BR feel that a computer-aided total train system is an achievable goal and a project team has been set up to investigate the possibility. From an energy conservation viewpoint, the advantage of computer based timetabling is that only those services that really need to run are allowed to do so. Energy savings result from the reduction in locomotive mileage.

Shipping, like airlines and railways, is a highly capital and energy intensive business, and similar benefits accrue from the use of scheduling packages. Actual details of the use of computers are scarce and only one case is known to have been reported [55].

## 2.5 Urban Traffic Control

Computer based urban traffic control schemes are now commonplace. The first scheme was implemented in Toronto in 1959 [83][84]. It was successful, and many other countries started developing their own schemes. The first British scheme was the West London Experiment which was inaugurated in 1968. Other cities followed on such as Leicester, Liverpool, Glasgow and Birmingham. Urban traffic control schemes are well documented [4][11][12] and a brief history is available from Honey [11].

Traffic control schemes are of benefit from many points of view. Most important of all as far as the planners are concerned is the reduction in traffic congestion that they bring. In the twelve months following the introduction of the West London Experiment, the area benefited from a 10% decrease in average journey times and surprisingly, a 14% reduction in the accident rate. Verification of the systems performance was provided in May 1975 when it became necessary to switch the system off for 48 hours - journey times increased by 30%. As far as this report is concerned, there are large benefits from an energy conservation point of view. It is estimated that the West London scheme paid for itself in the first six months in terms of decreased travel costs (including energy costs).

Once a system has been built, it is still possible to increase its effectiveness by refining the traffic control algorithms. A Glasgow experiment [16] investigated the effect of changing the algorithm of the computer controlling the traffic lights in the city centre. The algorithm that was originally in use was one that optimised the travelling time through the central area. This algorithm was superseded for the duration of the test (and afterwards), by an algorithm that optimised fuel consumption. The difference shown was that the new algorithm caused the car in the test to use 5.8% less fuel, whilst the average journey time increased by only 0.3%. A subsequent cost-benefit analysis showed that the new algorithm would save travellers through the central area an average of £44,000 pa.

Further advantages of traffic control systems are reductions in noise levels and pollution. A simulation study [17] has been undertaken in the USA that showed that pollution could be reduced by the appropriate design of urban traffic control systems.

## 2.6 Route Guidance Systems

Many kilometers are unnecessarily driven each year by drivers who do not choose the optimum route between their starting point and destination. Through lack of forewarning, they may also run into traffic congestion. Both of these things increase fuel consumption.

The results from a survey [85] of drivers route choice between towns in Gloucestershire were analysed to show the number of vehicle kilometers (vkm) that were unnecessarily driven through poor choice of route. The results were then extrapolated to cover the whole of the country and it was estimated that about 8300m vkm were wasted in 1976. The cost of this wastage was estimated to be about £530m. Of this, about £94m was the fuel cost.

Route guidance systems are a means for reducing this wastage. They are effective because they provide the driver with a stream of up to date information. This information is transmitted to the driver by either short distance radio transmitters placed in the vicinity of the junction, or by loop antennae situated under the road surface. A receiver within a vehicle approaching the junction decodes the information and presents it to the driver on a small display. This display will either show the route through the junction, or the actual road number and geographical direction.

Work by Jeffrey and Stockdale [29] has suggested that a typical British route guidance system could cost around £400m and return a net benefit of £1000m spread over 20 years. The cost to the motorist would be about £90 for the vehicle unit. An average motorist could expect to benefit to the tune of £14 pa. The average commercial user could expect to benefit by about £44 pa.

The economic viability of such systems is therefore in doubt, although experimental trials in Japan [36] seemed to indicate that these systems were economically attractive. Further problems may be encountered with commercial users because of the reluctant attitude of commercial drivers to any sort of electronic device in the cab.

The passage of time and continuing studies of route guidance systems may reveal that they are worth implementing on a large scale, perhaps only for reducing urban traffic congestion.

## 2.7 Conservation Through Telecommunication

Telecommunications may, in some circumstances, provide a satisfactory alternative for transport. Many types of passenger travel are undertaken for the interchange of information. The transfer of this information may be achieved with equal effectiveness by the use of some telecommunication or computer service. Similarly, many types of goods transport, such as the distribution of newspapers and mail are only forms of information transfer and they too are capable of substitution. Analysis has shown that telecommunication and computer services are very much less energy intensive than either personal or goods transport.

Telecommunications may reduce energy consumption in several less direct ways. For instance, the availability of cheaper and better telecommunications facility may result in changed patterns of personal behavior, home location and office location. It is not the purpose of this report to consider these indirect effects. The subject is well covered in the literature [13][14] to which the reader is referred.

### 2.7.1 Telecommunications as a Substitute for Personal Travel

A great deal of research and experimental work has been undertaken [13][14] in recent years upon the extent to which telecommunications can provide a satisfactory substitute for personal travel. A standard work on the subject by Lathey [86] reveals that some 16% of vehicle kilometers in the USA are potentially substitutable. The table on the following page shows how many vehicle miles are accounted for by various categories of travel.

**Table Two - The Potential Susceptibility of Travel to Telecommunications Substitution in the USA [86]**

| Category of travel | Vehicle miles (%) | Potentially substitutable (%) | Travel miles substitutable (%) |
|---|---|---|---|
| Transport of goods | 15 | 0* | 0* |
| Earning a living | 42 | 24 | 10 |
| Family business | 19 | 20 | 4 |
| Education/civic | 3 | 25 | 1 |
| Social/recreational | 21 | 5 | 1 |
| Totals | 100 | - | 16 |

*\* Transport of goods was not considered in this report.*

A substitutability figure is applied to each of these figures, and the total likely benefit in terms of vehicle miles saved is given.

Supportive results were obtained by a survey [87] of 6,400 face-to-face meetings conducted by the British civil service. The study revealed that only a third of those meetings were not capable of being replaced by some form of telecommunication service.

Many different forms of telecommunications service are already available for personal use. They range from the ubiquitous telephone through audio plus graphics, slow scan television up to full bandwidth conference television. The costs of these services varies quite considerably. Their benefit from an energy point of view varies to a similar extent. Lathey [86] estimates that a simple two-way telephone call has an energy advantage of about 500:1 compared with the average intercity business trip. Audio-visual conferencing is less advantageous with an energy benefit of about 11:1.

Several factors need to be taken into consideration when assessing the desirability of telecommunications substitution of transport. The convenience and accessibility of the service determines the enthusiasm with which the service is used to some extent. The use of the telephone poses few problems because of its familiarity and ease of use. Audio-visual conferencing is considered to be less convenient because of the need to travel to a studio and the need to take specially prepared material along. Psychological factors are also relevant because travel is often undertaken to maintain friendly business relations or to provide symbolic attendance.

Several American companies have been willing to try out some of the ideas reviewed in this section, and have been encouraged by the results achieved so far. NASA [14] reports that 17% of intersite travel has been replaced by an audio conferencing facility. The Dow Corporation [14] report a cost-benefit ratio of 2:1 in their similar system.

2.7.2 Telecommunications as a Substitute for Goods Transport

Letters, cheques, newspapers and books are examples of physical objects the purpose of which is to provide a convenient means of recording information which has to be conveyed through space and time. In electronic form however, that information could be transmitted without the transport of goods. There do appear to be some worthwhile possibilities for conserving energy in this way. The necessary technology - hardware and software - is in a state of very rapid development and the potential should become progressively greater over the next few years.

Postal services are major users of transport. In the UK for example, their fleet of 26,000 vehicles is one of the largest in the country. A US study using US Government data reached the conclusion that 13% of mail volume could be replaced by various forms of telecommunication. Using the same data but different assumptions, two other research studies obtained corresponding estimates of 31% and 65%. These figures seem to confirm that there is a large capability for reducing mail volume, and hence conserving energy.

The financial field provides examples to show how teleprocessing may come to reduce the demand for mail services. Banks are changing their basic mode of internal operation with the use of central computers which are readily accessible to branch offices through data links. Credit information is being retrieved quickly over telephone connections to computers. Centralized payroll accounting for large companies becomes possible because information can be disseminated by data links.

An experiment in Japan several years ago was mounted wherein edited newspapers were transmitted on spare CATV channels to facsimile receiving sets in private homes. Pages were sent as radio signals from the newspaper company in Tokyo to the experimental centre in Tama New Town. From there they were transmitted to individual households by coaxial cable. The receiving apparatus was controlled by a time switch which turned it on at certain times in the morning and evening. Such services could replace the daily delivery of newspapers with consequential savings in energy. If the final form were soft copy (i.e. a display on a TV set), rather than hard copy, additional savings would result from the reduced use of paper.

### 3.0 COMPUTERS. CONSERVATION AND INDUSTRY

Manufacturing industry uses about 40% of the total energy consumed in the United Kingdom [88]. Energy consumption per unit of output for industry as a whole has been decreasing steadily for many years. There are three reasons for this: the increasing share of output by the light industries; the replacement of coal-burning boilers by more efficient oil and gas-fired installations; and the introduction of new processes that have tended to favour energy efficiency, even though that was not the reason for their installation.

The difference in energy intensity of the different industries is considerable. They heavy industries, steel, glass, bricks etc are generally the most energy intensive. For this reason, they are the first to be considered in conservation studies [56], although their energy intensity will have ensured that efficiency has already received close attention.

Energy consumption in the heavy industries tends to be in large capital-intensive equipment with long lifetimes. Opportunities for conservation therefore, are likely to come from the fitting of energy management systems to existing equipment. Substantial savings are possible in most processes through improved instrumentation and control systems. Longer term measures in energy saving usually involve major investment decisions and, as with any investment considered by industry, will be looked at as a whole project, in financial terms, and in competition with a range of other possibilities for spending the available money. Energy conservation is only one of the factors taken into consideration in a project assessment and its relative importance will vary from industry to industry.

The computer has a fairly well defined role in industrial energy conservation. At the design stage, computers may be used to simulate the likely operational characteristics of several preliminary designs, and so help to select the final design. Process control is a well known and well documented computer application in the industrial field. Process control computers may be used to achieve some degree of plant energy optimisation. Other uses of the computer within the industrial environment are secondary to the main function of the plant. HVAC control and product distribution management are very much in evidence and can be energy optimised by computer, but since these computer applications are also relevant in other fields, they are treated separately elsewhere.

## 3.1 Simulation as a Design Tool

Simulation is a valuable tool for the reduction of energy consumption. By using a simulation model of the process at the design stage, it is possible to pinpoint design changes that will increase the energy efficiency. The main disadvantages of simulation are its complexity and high cost, Recently, simulation languages such as BDL[75] have been developed for specialised applications. They reduce the cost and length of time it takes to study particular applications. In the absence of a special language, a simulation study will be worthwhile if the project under investigation is highly capital intensive. Merette [66] describes a simulation study of a mill undergoing a major expansion. The conservation objective consisted of providing the management with a set of curves showing the production attainable as a function of the maximum allowed level of power demand, with different mill control strategies as parameters. Management were then able to decide upon the most effective course of action according to energy, economic, or other criteria.

Hoskins [67] describes a detailed computer model used to calculate the energy flows and electricity usage in domestic refrigerators and freezers. The model was then used to evaluate the energy and associated cost effects of various designs. The implementation of most of the measures revealed by the study showed that electricity consumption of a fridge could be reduced by 52% for a cost increase of only 19%. Domestic refrigerators use 1.5% of the UK's domestic energy [52] so on a national scale, results such as this are quite significant.

The use of simulation in the design of buildings is now a well developed practice in the USA. Several reports are available [73][74][75] which describe individual models of energy flows in buildings. These models generally predict the hourly demand patterns of the building from experimental evidence and then simulate the response of the heating, ventilation and air-conditioning equipment to these loads. The models will usually allow the determination of the effects of changes in building design variables on the energy consumption. Graven [75] describes an extensive model that even allows the incorporation of solar panels into the simulation.

Fehr et al [68] describe a project, the purpose of which was to produce a program and a questionnaire that could be used by householders to help reduce the heat loss in their homes economically. The householder fills in a simple form which includes questions about the essential characteristics of his home such as dimensions, number of windows, present insulation thickness, air conditioning and fuel type etc. The completed form is then input to the program, and the output obtained is an analysis of the home showing the economics of installing various types of energy conservation measures. This project was well received by the people who took part and motivated them to insulate their homes better.

## 3.2 Energy Management Systems

Energy management systems are becoming increasingly popular in industry. They are usually implemented as part of a plant-wide energy conservation drive. Manual conservation methods have their limitations in that the degree of control required by certain devices and processes is not possible manually. Computers provide the extra degree of refinement and can cut energy consumption drastically.

Computer equipment is widely available off the shelf from many of the large computer manufacturers, and from smaller firms who specialise in process control equipment. The cost of computer hardware can vary from about $5000 upwards. However, the cost of the hardware is not the complete story since sensors and actuators appropriate to the plant must be provided and the plant itself may require modification. Payback periods can range from a few months to a few years. The details of some success [60][69][70] stories can be found in the literature.

It is not the purpose of this report to describe in detail each of the many different application areas of energy management systems within an industrial environment. However, Table Three lists a small range of possible applications, many of which are described in the literature[56][64][63][76][77].

An energy management system will, in many cases, perform additional functions since most, if not all of the sensor equipment will already be available. Hamilton [63] gives an interesting example of this point, where an energy management system was installed on a boiler monitoring system as an afterthought. The original intention of, and justification for, the system was to provide only a monitoring and data logging function. The establishment benefited greatly from the energy management program.

Before an energy management system can be installed, it is necessary to perform an energy audit to ensure that the proposed system is justified on a cost basis alone, since this is the only acceptable method of assessing a systems worth.

**Table Three - Applications of Energy Management Systems and Related Functions**

Energy management
Demand control
Equipment scheduling
Temperature and humidity monitoring and control
Weather monitoring
Fan selection and speed control
Pump and valve monitoring and control
Refrigeration optimisation
Boiler and turbine control
Safety and security
Smoke alarm monitoring
Sprinkler monitoring
Exhaust fan emergency control
Flood alarm
Perimeter door control
Elevator control
Building operations
Equipment maintenance scheduling
Maintenance cost accounting

### 3.2.1 An Energy Survey

An initial step in the consideration of an energy management programme is the completion of an energy survey or audit. This follows the control engineers dictum "Monitor then Control". It is only through gaining a thorough understanding of a plants operation from an energy point of view that one can hope to make any savings. The energy survey will nearly always reveal a variety of very simple manual methods that will lead to an increase in energy efficiency, without recourse to computer methods.

Such a survey provides the data base necessary to plan a proper system installation. Electrical consumption and demand costs for the previous twelve months are needed, together with similar figures for other sources of energy. An inventory of energy consuming and producing devices is needed along with each of the devices characteristics such as load versus efficiency curves, maximum frequency of operation limits etc. Analysis of the survey data will reveal controllable loads and required monitoring points. Once operating strategies have been defined, and, given a typical pattern of energy usage, it is possible to calculate the energy savings resulting from the new strategy. At the same time, the cost of the monitoring and control system will have been worked out and the systems likely cost benefit ratio will be known.

System installation will usually follow a phased pattern, particularly if the system is large, so that experience can be gained as the system is expanded.

### 3.2.2 A Structured Approach

To ensure that an energy management system performs well, and that expansion plans are not limited by design, the system can be of a structured nature as proposed by Kaya [69].

Energy management systems can proceed at three increasingly complex levels of supervisions At the bottom level is the control of individual energy sources or sinks, like blast furnaces and turbines; At the intermediate level is group control of similar units used in parallel to ensure that group operation is optimal; Finally, at the highest level of control is a supervisory system for ensuring that the whole process is optimal.

Although primarily concerned with energy optimisation, a system of this nature must not prejudice other aspects of the plants operation.

At the lowest level of control, microprocessor supervision with appropriate sensors and actuators is sufficient to guarantee the best operating efficiency of, say, a boiler or turbine. For the higher levels, larger machines will be necessary. The sensors and actuators of the lower levels will be sufficient for these higher level functions. Despite the fact that each node in the network can be optimised with respect to its own energy efficiency, overall computer control is needed because optimum node efficiency does not guarantee total plant optimisation, and in some cases may actually impair it.

Consider the case of two boilers serving a single steam line. If they both have different non-linear load versus efficiency characteristics (because one is older than the other or uses a different fuel), then it is necessary to control them both as a pair to ensure optimum joint efficiency for a given steam load.

Figure Four shows how the different levels of control are related, and shows a division of a typical heavy industrial plant into power production equipment, power transmission equipment and power consumption equipment.

**Figure Four - Hierarchy of Computer Control**

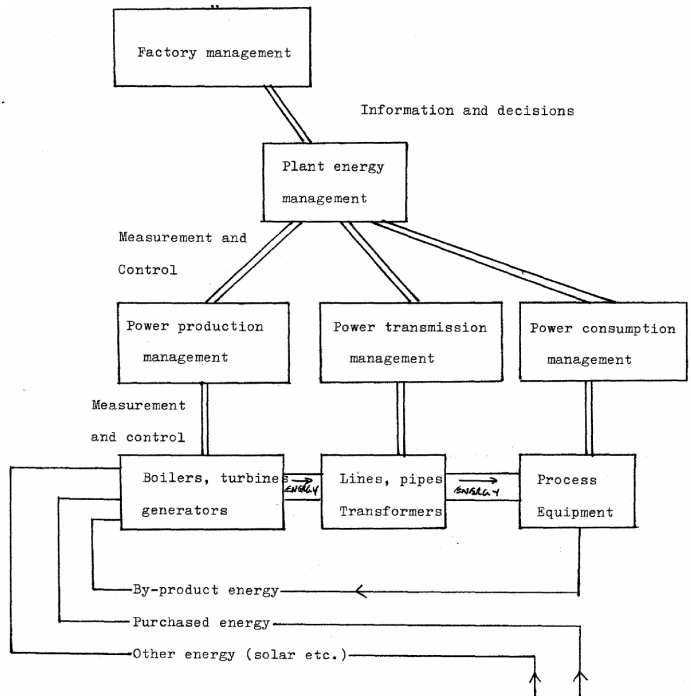

### 3.3 Demand Control

Commercial and industrial energy users are charged for electrical energy on the basis of two factors: the actual amount used and the peak demand reached for a specific length of time, usually 15 or 30 minutes. A demand control system continuously monitors the level of energy consumption, and, if peak is likely to occur, it starts shutting down non-critical items of equipment such as air-conditioning and electric motors. The demand charge imposed by the supply company theoretically covers the capital investment required to meet the collective peak demand of its customers.

Demand controllers only reduce total electricity consumption by a small amount, if at all. Their main purpose is to spread the load and so reduce the peak demand. On a large scale however, the use of demand controllers will increase the efficiency of electricity generation and distribution because the less efficient intermediate and peak demand equipment will not need to be used as often (see Figure Five). Electricity supply companies were not initially in favour of devices such as these because they reduced their revenue. This attitude is changing for the better because of the reduced need for expensive peak power generating capacity.

Demand controllers are in widespread use already. Several of the large computer manufacturers sell demand controllers, either as standalone systems, or as part of a more complex energy management system. There is a large amount of literature on the benefits of these systems [60]. Payback periods range from four months to a few years, depending upon exact circumstances, A group of three IBM system/7 demand controller users in the USA reported recently a saving of $1m in the first 12 months of operation.

Four different types of demand controller are currently available, each with a varying degree of complexity, effectiveness and cost:

1) Instantaneous kilowatts - Switches loads on or off depending upon the current demand. They have a tendency to switch loads off too soon and too often.

2) Average Demand - Switches equipment on or off according to the average demand over a predetermined previous period.

3) Dead band - Exhibits hysteresis and does not switch equipment back on until demand has fallen below a certain level. The dead band may be reduced with time so that there is less switching at the beginning of the demand period.

4) Forecasting demand - Switches devices on or off depending upon the predicted demand.

Before a demand system can be installed, it is necessary to investigate the electrical characteristics of the devices on a plant. This survey and engineering study should include as much information from plant operating personnel so that the demand control program takes into consideration any constraints applied to the system such as maximum or minimum on or off time of a given load and permissible start hour etc. Plant equipment can generally be divided into four categories according to the ease with which it may be computer controlled:

1) Inhibited - Loads which can be turned on by the computer.

2) Sheddable - Loads which are switchable entirely under program control.

3) Switchable - Loads which can be turned on or off by computer, but are subject to certain constraints.

4) Baseload - This equipment is not suited to computer control.

The amount of energy and money that can be saved in a demand control system depends to a large extent upon the number of devices that fall into the first three categories.

**Figure Five – Load Duration**

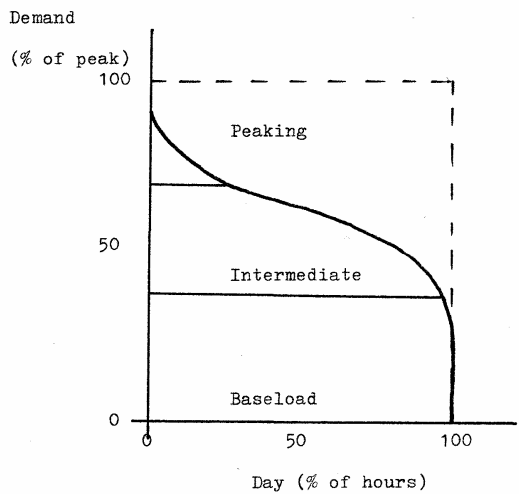

*The above curve shows the relationship between the amount of power that must be provided, and its duration. Most of the demand is satisfied by the baseload and intermediate generating systems, usually coal, oil and nuclear. The peak demand is catered for by gas-turbine and hydro systems. The baseload and intermediate capacity is more efficient than the peak capacity*

### 3.4 Electricity Supply Management

In order to operate economically and efficiently, an electricity supply company must match its instantaneous power generating capacity to the demand as closely as is possible. Some measure of demand forecasting is necessary because of the length of time it takes to bring a generating station up to full capacity. Electricity suppliers, particularly in those countries where generating capacity is privately owned, must also manage the sale and purchase of electricity between neighboring companies.

Computer technology is increasingly being used to help perform these functions more efficiently. Electricity supply networks are invariably large and several computers will be used, typically in some form of structured control hierarchy.

The principal function of the lower region of the control structure is to perform automatic generator control in real time. This combines load and frequency control with economic plant scheduling, The computer continuously determines generating reserves and, if preset limits are reached, produces and alarm. in the event of an alarm, an emergency routine automatically bypasses economic considerations and resolves the conflict.

At the top end of the control structure, computers can help to manage a wide variety of contracts between electricity supply companies for the sale or purchase of electricity. This may be required to cope with emergencies, to cope with maintenance schedules, or to improve load factors.

In the day-to-day operation of the plant, the peak load is estimated according to weather forecasts. Then predicted hourly loads are calculated for the next few days from known load patterns and seasonal demand records. The contracted power interchanges between companies are also calculated for the immediate future. From this information, the computers can then schedule the various generating and switching centres to give the most economic and energy efficient set up for the next 24 hours.

The use of the computer in electricity supply management brings about certain improvements in overall system efficiency and reduces the probability of total network failure such as happened in New York in 1965. Tyler [14] estimates that improving the efficiency of electricity supply and distribution from 25% to 30% will bring about a 5% reduction in UK primary energy consumption.

The computer is by no means the only possible means of improving efficiency in this context but it is certain that the increased use of computers will have a bearing upon the efficiency of electricity production.

### 4.0 CONSERVATION IN THE COMMERCIAL SECTOR

For the purposes of this report, the commercial sector is defined to be the office environment in which most non-manual workers work. The opportunities for computer-aided energy conservation in this sector are limited to two things.

Firstly, computers may be used to help design buildings so that they require less energy to keep the occupants warm. The results from design studies are very promising as mentioned in section 3.1, however, the method as a whole is limited by the very low rate at which buildings are replaced or built, compared to the number of buildings that actually exist.

Secondly, computers may be used in heating, ventilation and air-conditioning control systems. The number of possible sites that could use a computer based HVAC control system is enormous. The systems are relatively easy to install and few buildings have any form of HVAC control already.

A major obstacle to the more widespread implementation of HVAC control systems is the general lack of awareness as to the potential benefits of these systems. Any efforts made towards increasing awareness should bring about a significant reduction in the amount of energy consumed in the commercial sector.

### 4.1 Heating, Ventilation, Air-conditioning & Control Systems

The control of HVAC systems is fundamentally a part of the subject of energy management in general. The reasons for treating HVAC systems separately are that they are a well-defined sub-problem and that the potential market is very large.

HVAC controllers range in complexity from simple 16 channel microprocessor based devices at one extreme to midi and maxi machines with thousands of channels at the other. The market tends to be dominated by smaller systems because of their low cost and ease of implementation. The installation of a HVAC control system does not usually pose any great problems provided that an extensive energy audit has been performed so that areas of maximum benefit are well known.

An energy audit will usually identify the devices that are open to control, and a system control strategy can be developed according to knowledge of the behavior of the buildings occupants. Typically, all but the smallest systems will be aware of the usual working hours, lunch-breaks, weekends and works holidays. This information can then be used to schedule the various devices that comprise the HVAC system so that conditions of adequate comfort are maintained whenever the building is occupied.

The simpler HVAC controllers may be limited in their control strategies to providing little more than temperature based on/off switching for the heat-producing devices. Medium to large scale controllers will monitor the following types of device: heating equipment, fans, chillers, air ducts, ventilation openings, humidifiers and water heaters. The controller will schedule these devices so that they do not work against each other (often the case with mechanical systems), and that they work in such a fashion to provide the right working environment for the least energy consumption. The more complex HVAC systems often perform functions in addition to the above. These systems, often known as Building Automated Systems (BAS) will control such things as the buildings security and access, lifts, escalators, fire alarms and lighting.

In large buildings, distributed systems are used. Remote devices perform a data-logging and multiplexing function for several adjacent items of plant. The remote devices only pass information to the main machine when requested to do so, or when an input changes state. Periodically each remote station will perform its own integrity check.

The potential benefits of computer based HVAC controllers are very large indeed. They are much more intelligent than previous systems and their inherent flexibility means that they can be matched to the precise requirements of the user. Self adapting systems are available that are capable of learning the thermal characteristics of the building and matching themselves to that behavior.

It is very important that companies are made more aware of the benefits of better HVAC control systems because space heating consumes so much of the worlds energy and current systems are, more often than not, inefficient.

## 5.0 CONSERVATION IN THE DOMESTIC SECTOR

The UK energy statistics for 1978[88] show that some 26% of the total energy consumed on a heat supplied basis was used by the domestic sector. The breakdown of the energy consumed is: space heating 66% water heating 21% and lighting and miscellaneous 3%. Thus the major part of domestic energy is used for maintaining thermal comfort. Studies have shown that people from all over the world prefer to live and work within certain temperature bands. The idealistic temperature pattern over the day varies from 13-17°C during housework and up to 20-23°C during relaxation in the evenings. Since people have these fairly wide comfort bands, it follows that their motivation to change the temperature whilst it remains within these bands will be low. Since every extra degree centigrade of indoor temperature accounts for a further 5%-10% of space heat consumption, the maintenance of the temperature towards the lower edge of these bands is in the national interest.

The efficiency of the heating control system affects to a certain extent the choice of temperature. For example, if a simple warm air heating system has a thermostat that switches the heating either on or off, then the occupants will tend to keep the average temperature rather higher than is necessary to ensure that the room temperature never falls below a certain minimum. If people are to change their temperatures, then it is desirable to have a simple and responsive control system.

The advent of the microprocessor has opened up new possibilities for central heating control. Microprocessor based central heating controllers can be as responsive as the heat production equipment allows and can be designed to keep certain rooms at a particular temperature. The microprocessor will also permit the implementation of more complex sequencing commands so that, for example the heating program can be different at weekends.

There are two problems that may hinder the acceptance of microprocessor based central heating controllers.

Firstly, there are problems with the price of such equipment. Although a microprocessor system can be expected to improve the thermal efficiency of a home, their initial cost could possibly outweigh the economic advantages of their use. The expense arises not from the microprocessor, but from the sensing and actuating equipment that is needed. Present day systems do not require this equipment because, except for one room, they use open loop control which does not require any remotely operated valves.

Secondly, more responsive central heating controllers have man/machine interaction problems. It is difficult to design a simple, efficient and cheap interface on the controller that is capable of supporting the advanced dialogue that a microprocessor would allow. Although microprocessor based central heating controllers seem to be an easy way of conserving energy in the home, it is unlikely that they will be accepted until the problems of price and. interface design have been solved. A much more cost-effective way to save energy in the home is to insulate the roof, walls and windows, and to switch off unnecessary lighting.

## 6.0 CONCLUSIONS

This report has described in detail each of the many different ways that computer technology may be used to conserve energy.

The cost of computer technology is decreasing rapidly and is helping the cause of conservation by bringing computer control economically to increasingly marginal applications.

Computers are not being used to the fullest extent however, because of a general lack of awareness of the possible computer applications and the benefits that they can bring. Energy will never be as cheap or as readily available as it has been in the past, and so it is important that the realisation of the potential benefits of computer methods occurs as soon as is possible.

In conclusion therefore, computer based energy conservation is economic in many applications and it is important that this is more widely recognised so that coordinated efforts can be made to conserve energy on a large scale.

### 6.1 Suggestions for Further Work

This report has not, in general, given exact figures for the likely effect that the implementation of computers for energy conservation purposes would have upon national energy consumption figures. Further work could recover figures of this sort from the literature and use them to estimate the current potential of computer based energy conservation-measures. This potential, probably expressed as a percentage saving over current consumption, could then be used to encourage the more widespread implementation of computer based energy conservation.

## ACKNOWLEDGMENTS

I would like to thank IBM United Kingdom Laboratories Limited for their generous help in providing the initial literature searches and documents upon which this project is based. I would also like to thank my father, Mr. W.J.S. Croll, my Tutor and Supervisor Mr. A.D. Elliman and Mr. A.A. Hamilton for their invaluable advice and assistance during the course of this project.

77) Energy Management by Computer Control at Uniroyal Rubber Factories. Craemer R.H. et al. IEEE Publ.77CH1184-1 1977.

82) Computer Scheduling of Vehicles From One or More Depots to a Number of Delivery Points. Holliday A., and Wren A. Op Res Qtrly Vol 23(3).

83) The Metropolitan Toronto Signal System. Hewton J.T. Joint Symp. on Area Traffic Control of Road Traffic. Inst. Civil Engineers London Feb 1967.

84) Toronto's Digital Computer Signal System. Traffic Eng. and Control. Vol 12(9) Jan 71.

85) Inter Urban Route Choice Study - Driver Interviews and Journey Time Survey in Gloucs. A Report Prepared for the DOE ECEF Div by Transport Planning Associates May 77.

86) Telecomms. Substitutability for Travel: an Energy Conservation Potential. Office of Telecomms. US. Dept Commerce 75, Lathey C.E.

87) Reducing the Need for Travel. Kryzezkowski R. and Hennemen S., Dept. Transp. Urban Mass Transportation Div Mar 74.

88) UK Energy Statistics 1978. HMSO.